\documentstyle[twoside,fleqn,espcrc2,epsf]{article}

\newcommand{\AmS}{{\protect\the\textfont2
  A\kern-.1667em\lower.5ex\hbox{M}\kern-.125emS}}

\newcommand{\lesssim}{ {\
\lower-1.2pt\vbox{\hbox{\rlap{$<$}\lower5pt\vbox{\hbox{$\sim$}}}}\ }  }
\newcommand{\gtrsim}{ {\
\lower-1.2pt\vbox{\hbox{\rlap{$>$}\lower5pt\vbox{\hbox{$\sim$}}}}\ }  }

\def\la{\langle}
\def\ra{\rangle}
\def\l{\left}
\def\r{\right}
\def\lqcd{\Lambda_{QCD}}

\def\plb#1{Phys.~Lett.~{\bf B#1}}
\def\npb#1{Nucl.~Phys.~{\bf B#1}}
\def\prl#1{Phys.~Rev.~Lett.~{\bf #1}}
\def\prd#1{Phys.~Rev.~{\bf D#1}}
\def\zpc#1{Z.~Phys.~{\bf C#1}}

\newdimen\unit
\def\point#1 #2 #3{\vbox to0pt{\kern-#2\unit
\hbox{\kern#1\unit#3}\vss}
 \nointerlineskip}
\title{How Small Can the Light Quark Masses Be?}

\author{Laurent Lellouch\address{Centre de Physique Th\'eorique, 
CNRS-Luminy, Case 907,
F-13288 Marseille Cedex 9, France}
        \thanks{Many thanks to my collaborators
E. de Rafael and J. Taron~\protect\cite{LLeRT97}. CPT is UPR 7061.}}
       
\begin{document}

\begin{abstract}
We derive rigorous lower bounds for the combinations of light 
quark masses $m_s+m_u$ and $m_d +m_u$. 
\end{abstract}

\maketitle

\vspace*{-50mm}
\hspace*{135mm}
%\begin{flushright}
\makebox[2cm]{CPT--97/P.3532}
%\end{flushright}  
\vspace*{33mm}

\section{Introduction}
\vspace{-0.2cm}
The masses of the light $u$, $d$ and $s$ quarks are fundamental
parameters of the Standard Model and play an important role in low
energy hadron phenomenology as well as in our understanding of both
chiral symmetry breaking and the origin of mass. Until perhaps a year
ago, a certain concensus had been reached as QCD
sumrules~\cite{JBiPR95} and quenched lattice QCD~\cite{CAletal94}
yielded compatible values.  However, recent analyses of the cutoff
dependence of quenched lattice results have yielded significantly
lower values, with an indication that unquenched values could be even
smaller~\cite{RGuB97,BGoetal97}.  Perhaps even more surprising are the
results of the dynamical Wilson fermion calculation by
SESAM~\cite{NEietal97} who find very small $u$ and $d$ masses but a
strange quark mass in agreement with the QCD sumrule determinations.
(For even more recent developments, see~\cite{RGulat97}.)  In light of
this rather confusing situation, we have sought to derive rigorous
lower bounds for the combinations of light quark masses $m_s +m_u$ and
$m_d +m_u$ below which lattice results cannot fall without implying
a surprising breakdown of perturbative QCD.

In our analysis, we consider two channels. The first,
corresponding to the two--point function
\begin{equation}
\psi_5(q^2) = i\int d^4x\ e^{iq\cdot x}\la 0|T\l\{A(x)
A^{\dagger}(0)\r\}|0\ra
\ ,
\end{equation}
with $A\equiv\partial_\mu A^\mu{=}i(m_s+m_u){:}\bar s\gamma^5 u{:}$ or 
$i(m_d+m_u){:}\bar d\gamma^5 u{:}$, is a {\bf pseudoscalar}
channel. The second is a {\bf scalar-isoscalar} channel and is 
associated with
\begin{equation}
\psi(q^2) = i\int d^4x\ e^{iq\cdot x}\la 0|T\l\{S(x)
S^\dagger(0)\r\}|0\ra
\ ,
\end{equation}
where $S{=}\hat m [{:}\bar uu{:}+{:}\bar d d{:}]$ 
and $\hat m{=}(m_u+m_d)/2$.

Both two--point functions satisfy a twice subtracted dispersion
relation,
\begin{equation}
\psi''_{(5)}(q^2) = 
\frac{2}{\pi}\int_0^\infty dt \,\,\frac{\mbox{Im}\psi_{(5)}(t)}
{(t-q^2)^3}
\ ,\label{eq:disprel}
\end{equation}
which relates an integral of the hadronic spectral functions,
$\mbox{Im}\psi_{(5)}$, to the QCD behavior of the two--point functions in
the deep euclidean region. 

For $-q^2{=}Q^2{\gg}\lqcd^2$, the LHS has an OPE,
\begin{equation}
\psi''_{(5)} {=} \frac{m^2}{Q^2}\l\{\psi_{(5)0}'' {+} \psi_{(5)2}''
\frac{1}{Q^2}
{+}\psi_{(5)4}''\frac{1}{Q^4}+\cdots\r\}
,
\label{eq:qcd2pt}
\end{equation}
where $\ m\ $ is\  the\  relevant\  mass\ \ combination, the 
$\psi_{(5)0}''$ are
known to four loops~\cite{KCh97}, and $\psi_{(5)2}''$, 
$\psi_{(5)4}''$ contain mass and condensate corrections~\cite{npert}.

The RHS is determined by the intermediate hadronic states which can
contribute to the spectral functions through unitarity:
\begin{eqnarray}
&&\frac{1}{\pi}\mbox{Im}\Psi_{(5)}(q^2)=
(2\pi)^3\ 
\sum_{\Gamma}\,\delta^{(4)}(q-\sum p_{\Gamma})\,\times\nonumber\\
&&\qquad\times\la0\vert
J(0)\vert\Gamma
\ra\la\Gamma\vert J^\dagger(0)\vert 0\ra
\ ,
\label{eq:spectral}
\end{eqnarray}
with $J{=}A,S$. 
For the {\bf pseudoscalar} case, the lowest--mass contribution is 
$|\Gamma\ra{=} |K^+\ra$ (or $|\pi^+\ra$ for the $\bar d\gamma^5u$
current) while for the {\bf scalar} case it is $|\pi\pi\ra$. 
The $K^+(\pi^+)$ contribution is determined by $f_{K(\pi)}$ and
$M_{K(\pi)}$. The
$|\pi\pi\ra$ contribution is determined by a single form factor $F(t)$,
which has been well studied in the low--energy region~\cite{JDoGL90}.

It is clear from Eq.~(\ref{eq:spectral}) that the spectral functions
are sums of positive terms so that keeping only the lowest--mass contributions
gives lower bounds to these functions. Then, Eq.~(\ref{eq:disprel})
and the fact that the $\psi_{(5)}''$ are proportional to the square of
the quark masses of interest, can be used to translate 
this bound into a lower bound on these masses. 
%It must be emphasized
%that this derivation is based on first principles.

\section{Pseudoscalar Channel}
\vspace{-0.2cm}
Following \cite{CBeNRY81} 
we consider moments of the hadronic continuum integral
{\small ($t_0{=}(M_{K,\pi}{+}2M_\pi)^2$)}:
\begin{eqnarray}
&\Sigma_N(Q^2) \sim (t_0+Q^2)^N
\l(\frac{\partial}{\partial Q^2}\r)^N\l(\psi_5''(Q^2)-\r.\nonumber\\
&\qquad-\l.K^+(\pi^+)\ \mbox{contribution}\r)
\ .
\label{eq:contin}
\end{eqnarray}
Positivity of the spectral function then implies, for the first
three moments,
\begin{equation}
\Sigma_N\ge 0, \quad \Sigma_{M}\ge\Sigma_{M+1}, \quad
\Sigma_0\Sigma_2\ge (\Sigma_1)^2
\ ,
\end{equation}
with $N{=}0,1,2$ and $M{=}0,1$.

\begin{figure}[tb]
\setlength{\epsfxsize}{100mm}
\vspace{-1cm}
\centerline{\epsfbox{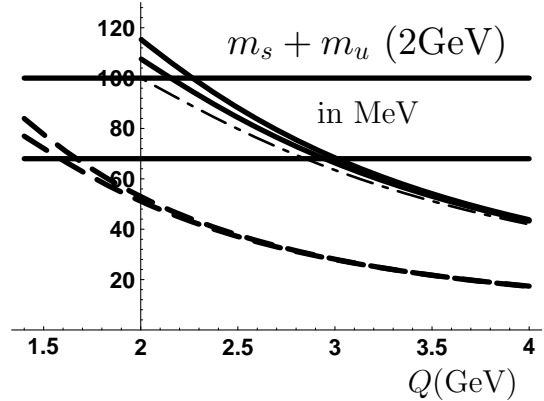}}
\vspace{-8cm}
\unit=0.8\hsize
\point 0.90 0.06 {\large{$Q(\mbox{GeV})$}}
\point 0.50 0.82 {\Large{$m_s+m_u\ (2\mbox{GeV})$}}
\point 0.70 0.65 {\large{in MeV}}
\vspace{-0.7cm}
\caption{\protect\small Bounds from $\Sigma_0{\ge}0$ (dashed) and 
quadratic bounds (solid) for $\Lambda_{\overline{MS}}^{(3)}\,{=}290,380$ MeV. 
The dot-dashed curve illustrates the uncertainty due to unknown
${\cal O}(\alpha_s^4)$ terms 
($\Lambda_{\overline{MS}}^{(3)}\,{=}380$ MeV). Values of $m_s+m_u$
below the various bounds are excluded. 
The horizontal lines correspond to the central quenched (upper) and
unquenched (lower) lattice values of \protect\cite{RGuB97}. }
\vspace{-0.7cm}
\label{fig:ms}
\end{figure}

The first bound we consider is the one arising from $\Sigma_0{\ge} 0$.
We find, for the running quark masses:
\begin{eqnarray}
&\left[m_{s}(Q^2)+m_{u}(Q^2)\right]^2 \ge \frac{16\pi^2}{N_c}
\frac{2f_{K}^2 M_{K}^4}{Q^4}\times
\label{eq:sig0bnd}\\
&\times\frac{1}{\left(1+\frac{M_{K}^2}{Q^2}\right)^3}
\frac{1}{\left[1+
\frac{11}{3}\frac{\alpha_{s}(Q^2)}{\pi} +\cdots\right]}
\ ,\nonumber
\end{eqnarray}
where $Q^2$ is the scale at which the OPE of Eq.~(\ref{eq:qcd2pt})
is performed. Using three--loop running, we shall give
all numerical results
in the $\overline{MS}$ scheme at a reference scale of $4\mbox{ GeV}^2$.
Clearly, this bound falls off quickly with $Q^2$, 
so one would like to work with as small a $Q^2$ as possible. However
at small $Q^2$, the OPE of Eq.~(\ref{eq:qcd2pt}), and in particular
the pQCD expansion, breaks down. We consider that values
of $Q{\gtrsim} 1.4\mbox{ GeV}$ are safe. Our results are shown
in Fig.~\ref{fig:ms}. 
For low $Q$, the preliminary, central, 
unquenched results of \cite{RGuB97,BGoetal97} are excluded. 
It should be emphasized
that in Fig.~\ref{fig:ms} we are comparing a ``calculation''
of the quark masses with a bound which is saturated in the extreme
case where all contributions to the spectral function
other than $K^+$ are neglected.

The bounds arising from $\Sigma_{1,2}{\ge} 0$ give an 
improvement due to the fact that higher and higher moments increase
the weight of the low--energy part of the hadronic spectral function.  
Some of this improvement is offset by the fact that the OPE converges
less and less well so that one has to go out to larger values of $Q$.
A similar phenomenon occurs when one considers the quadratic bound
given by $\Sigma_0\Sigma_2{\ge} (\Sigma_1)^2$.  Nevertheless, the net
improvement is still significant, as seen in
Fig.~\ref{fig:ms}, where we believe that values of $Q{\gtrsim} 2\mbox{
GeV}$ are safe.  We
find that even the central quenched values of \cite{RGuB97,BGoetal97} 
may be excluded.
This bound allows less trivial continuum spectral functions than the bound of
Eq.~\ref{eq:sig0bnd}:
a $\delta$-function with arbitrary position and
weight saturates it.

\begin{figure}[tb]
\setlength{\epsfxsize}{100mm}
\vspace{-1cm}
\centerline{\epsfbox{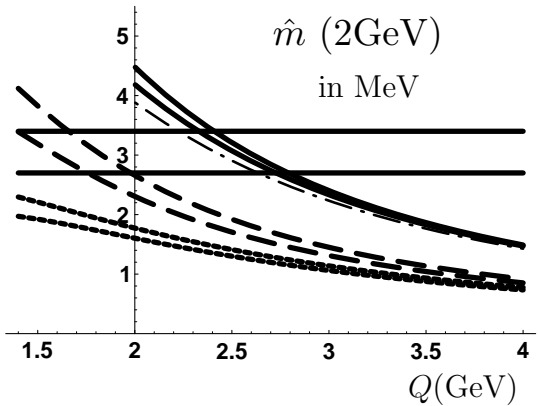}}
\vspace{-8cm}
\unit=0.8\hsize
\point 0.90 0.06 {\large{$Q(\mbox{GeV})$}}
\point 0.60 0.85 {\Large{$\hat m\ (2\mbox{GeV})$}}
\point 0.70 0.72 {\large{in MeV}}
\vspace{-0.7cm}
\caption{\protect\small Quadratic bounds (solid) for 
$\Lambda_{\overline{MS}}^{(3)}\,{=}290$, 380 MeV. 
The dot-dashed curve illustrates the uncertainty due to unknown
${\cal O}(\alpha_s^4)$ terms 
($\Lambda_{\overline{MS}}^{(3)}\,{=}380$ MeV). The long-dashed
and short-dashed curves correspond, respectively, to the scalar bounds with and
without information on $F$'s
phase. The 2 curves in each set reflect variations
of all input parameters, including $\Lambda_{\overline{MS}}^{(3)}$,
in their error ranges. Values of $\hat m$
below the various bounds are excluded. 
The horizontal lines correspond to the central quenched (upper) and
unquenched (lower) lattice values of \protect\cite{RGuB97}.}
\vspace{-0.7cm}
\label{fig:mumd}
\end{figure}

A quadratic bound can also be derived for $\hat m$ and is shown in
Fig.~\ref{fig:mumd}. Again, the lattice results of 
\cite{RGuB97,BGoetal97,NEietal97}
are in serious difficulty. Note that this bound 
is significantly more stringent than those
of \cite{TBhGM97} where different conclusions were drawn.

The known radiative corrections to the $\Sigma_N$ are large (more so
as $N$ increases) and positive, and exhibit near geometric growth.  To give
some measure of the uncertainty associated with potentially large,
higher--order terms, we also show in Figs.~\ref{fig:ms} and
\ref{fig:mumd} the quadratic bounds obtained by assuming that the
${\cal O}(\alpha_s^4)$ terms are equal to the ${\cal O}(\alpha_s^3)$
terms to the power 4/3. Should this behavior persist
to even higher orders, the bounds would be lowered further.

\section{Scalar-Isoscalar Channel}
\vspace{-0.2cm}
As noted above, the lowest mass contribution to the scalar-isoscalar
channel is determined by a form factor, $F(t)$. $\chi PT$ to two loops
and a dispersive analysis yield~\cite{JDoGL90}:
\begin{equation}
F(t) = F(0)\l[1+(\la r^2\ra^\pi_s/6)\, t+{\cal O}(t^2)\r]
\end{equation}
with
$F(0){=}M_\pi^2(0.99{\pm}0.02){+}{\cal O}(M_\pi^6)$ and $\la r^2\ra^\pi_s
= (0.59{\pm}0.02)
\mbox{ fm}^2$. Following the methods of \cite{SOkS71}, we 
construct bounds on $\hat m$ using this information. The QCD behavior
of $\psi''$ is similar to that of $\psi_5''$ up to very small
$m_{u,d}$ mass corrections which we neglect. The corresponding
OPE appears well behaved for $Q{\gtrsim} 1.4\mbox{ GeV}$. The bound we obtain
is
\begin{eqnarray}
&\hat m^2(Q^2)\ge 
\frac{\pi}{N_c}\frac{M_{\pi}^6}{Q^4}
\frac{12\times (0.99\pm 0.02)^{2}}
{\left(\sqrt{1+\frac{4M_{\pi}^2}{Q^2}}+\frac{2M_{\pi}}{Q} \right)^6}
\times\nonumber\\
&\times\frac{1+|{\cal F}\l(Q^2,M_\pi^2,\langle r^{2}\rangle_{s}^{\pi}\r)|^2}
{\left[1+\frac{11}{3}\frac{\alpha_{s}(Q^2)}{\pi}+\cdots
\right]}
\ ,
\end{eqnarray}
where ${\cal F}$ is a nontrivial function of its arguments. Though
this bound is quite similar to the axial current bound of 
Eq.~(\ref{eq:sig0bnd}), it is $1/N_c$-suppressed and chirally
suppressed by an additional 
factor of $M_\pi^2$. One might expect, therefore, that it
would be much worse; yet it is surprisingly competitive (see
Fig.~\ref{fig:mumd}). 
This
is due to the fact that the absence of a prominant scalar-isoscalar
resonance is compensated by rather large chiral loops and the fact
that, numerically, $M_\pi$ and $f_\pi$ are not so different.

This bound can be improved by taking into account the phase of
$F(t)$ which is given by the $I{=}J{=}0$, $\pi$--$\pi$ phase shift in
the elastic region $4 M_\pi^2{\ge} t{\ge} 16 M_\pi^2$. Incorporation
of this information requires solving two
integral equations~\cite{MMi73}.
Restricting to a conservative region, $4 M_\pi^2{\ge} t{\ge} 
(500\mbox{ MeV})^2$, 
we use the phase-shift given by a resummation of chiral logs calculated
at the two--loop level, which agrees well with available 
experimental information~\cite{JDoGL90}.
The improved bound we obtain is shown in Fig.~\ref{fig:mumd}. 
Again, we find
that the lattice determinations of \cite{RGuB97,BGoetal97,NEietal97} 
are in difficulty with
this bound.
We should add that the sumrule results of 
\cite{JBiPR95} are all compatible
with our lower bounds. Also, since this talk was given, two other
attempts at putting bounds on the light-quark masses 
have been reported \cite{FYn97,HDoN97}.

\end{document}